\documentclass[a4paper]{spie} 

\usepackage{amsmath,amsfonts,amssymb}
\usepackage{graphicx}
\usepackage[colorlinks=true, allcolors=blue]{hyperref}
\usepackage{textcomp}

\title{LINC-NIRVANA Commissioning at the Large Binocular Telescope - Lessons Learned}

\author[a,e]{Kalyan Kumar Radhakrishnan Santhakumari}
\author[a,e]{Carmelo Arcidiacono}
\author[a,e]{Maria Bergomi}
\author[b]{Thomas Bertram}
\author[b]{Florian Briegel}
\author[a,e]{Jacopo Farinato}
\author[b]{Thomas M. Herbst}
\author[c]{John M. Hill}
\author[b]{Micah Klettke}
\author[a,e]{Luca Marafatto}
\author[d]{Rosalie C. McGurk}
\author[a,e]{Roberto Ragazzoni}
\author[b]{Fabio P. Santos}
\author[a,e]{Valentina Viotto}
\affil[a]{INAF-Osservatorio Astronomico di Padova, Vicolo dell'Osservatorio 5, 35122 Padova, Italy}
\affil[b]{Max Planck Institute for Astronomy, K\"onigstuhl 17, 69117 Heidelberg, Germany}
\affil[c]{Large Binocular Telescope Observatory, 933 N Cherry Ave, Tucson, AZ 85721, USA}
\affil[d]{Carnegie Observatories, Pasadena, CA 91101, United States}
\affil[e]{ADONI, Laboratorio Nazionale di Ottica Adattiva, Italy}

\authorinfo{Send correspondence to Kalyan Radhakrishnan: \href{kalyan.radhakrishnan@inaf.it}{kalyan.radhakrishnan@inaf.it} }

\begin{document} 
\maketitle

\begin{abstract}
LINC-NIRVANA (LN) is one of the instruments on-board the Large Binocular Telescope (LBT). LN is a high-resolution, near-infrared imager equipped with an advanced adaptive optics module. LN implements layer-oriented Multi-Conjugate Adaptive Optics (MCAO) approach using two independent wavefront sensors per side of the binocular telescope measuring the turbulence volume above the telescope. The capability of acquiring up to 20 Natural Guide Stars simultaneously from two distinct fields of view, and using them for wavefront sensing with 20 separate pyramids per side of the telescope makes the LN MCAO system one of a kind. 

Commissioning of the left MCAO channel is almost complete, while that of the right arm is on-going. The Science Verification on the left side is expected to start soon after the MCAO performance is optimised for faint guide stars. In this article, we put together the lessons learned during the commissioning of the LN MCAO module. We hope and believe that this article will help the future MCAO instrument commissioning teams.
\end{abstract}

\keywords{adaptive optics, MCAO, LINC-NIRVANA, commissioning, lessons-learned, LBT}

\section{INTRODUCTION}
LINC-NIRVANA\footnote{\url{http://www.mpia.de/LINC/}} (LN) is a high-resolution, near-infrared imager mounted at the rear, bent-Gregorian foci of the Large Binocular Telescope (LBT) \cite{2019LNworkingPaperInPrep, 2018SPIEHerbstInstallation}. LN has a unique Multi-Conjugate Adaptive Optics (MCAO) module that uses a layer-oriented approach\cite{2018SPIEHerbstMCAO}. Two independent wavefront sensors per side of the binocular telescope measure the turbulence volume above the telescope. One is conjugated to the ground layer, called the Ground-layer Wavefront Sensor (GWS), and the other to a higher altitude, called the High-layer Wavefront Sensor (HWS). Both GWS and HWS use Natural Guide Stars (NGSs) from two adjoining but distinct fields of view. While the GWS can acquire up to 12 stars from the 2'-6' diameter Field of View (FoV), the HWS can acquire up to 8 stars from the inner 2' diameter FoV. The footprints of the stars overlap one over the other at the ground-layer. However, at the high-altitude conjugated layer, the stars' footprints are spatially separated, depending on the asterism. Therefore, the HWS has to operate under partially illuminated conditions. We have a simple but robust solution for this issue\cite{JATIS2019Kalyan}. Altogether, the MCAO correction is expected to provide uniform, diffraction-limited Point Spread Function (PSF) throughout the science camera. The only other working nighttime MCAO system in the world is the GeMS at the Gemini South telescope\cite{GeMsMCAO1, GeMsMCAO2}.

The LN team, with substantial support from the LBT mountain crew, installed the fully aligned and integrated LN on the LBT central instrument platform in September 2016. The details and specifics about the LN MCAO system and its AIV are all already existing in conference proceedings \cite{2018SPIEHerbstInstallation, 2018SPIEHerbstMCAO, 2015aoelLUCA, 2014SPIEKalyan, 2014SPIEMaria}. We encourage interested readers to look into those. After a pair of pre-commissioning runs that focused on the alignment of bulk optics of the telescope to the instrument and calibration of one of the GWSs, the commissioning of the left arm of the MCAO system began on March 2017, and on March 2018 we declared our First Light MCAO for LN\cite{2018SPIEHerbstInstallation, 2018SPIEHerbstMCAO}.

In this paper, we outline the lessons learned during the commissioning of the LN MCAO module. The lessons learned span various regimes. For example, (1)~the low-frequency vibrations seen by the system, (2)~AO calibration updates, (3)~temperature variations at LBT, (4)~efficient acquisition of the multi-pyramid system, (5)~flexure tracking, (6)~strategic planning and implementation of half-night, full-night, and daytime commissioning activities across the year, etc.

The LN MCAO system may be considered a test-bed for future ELT-based MCAO systems, at least in terms of complexity, and the lessons learned during commissioning provide valuable insight into the design and development of such systems. Through this article, we aim to transcribe our commissioning experience for the benefit of future MCAO instrument commissioning teams.

\section{Vibration Environment at the LBT}
The 20\textsuperscript{th} century witnessed an increase in the number of optical-IR ground-based telescopes. Telescope size ranged from 1\hspace{1pt}m to 10\hspace{1pt}m. Hereafter, by telescope we mean optical-IR ground-based telescope with size larger than 1\hspace{1pt}m and used for nighttime astronomy. The astronomy community had to wait until the 1960s to make use of the full capabilities of the telescope in real-time i.e., reaching the diffraction limit of the telescope using AO. Before the implementation of AO in telescopes, science images were dominated by aberrations due to the atmospheric turbulence. When AO systems started to provide almost flat-wavefronts, removing the effects of the atmosphere in real-time, vibrations became significantly important. 

Today, vibrations at the telescope are a significant contributor to the error budget of the residual wavefront error. Some science cases require to have an almost vibrationless system. Two explicit examples are (1)~observations that require high-Strehl, high-contrast, and high-resolution abilities (for example, exoplanet imaging) and (2)~observations using interferometric instruments where the stability of the measured fringes is directly related to the strength of the vibrations. 

The primary sources of mechanical vibrations in the telescope are cryo-coolers, pumps, fans, and motors. The noise injected by these components can excite the resonant and harmonic frequencies of the other mechanical components within the system. Then there are vibrations introduced by the wind. For example, at the LBT, looking into a 15\hspace{1pt}m/s wind makes the closed loop difficult or impossible to manage\cite{2016SPIEHill}. The impact of vibrations can translate to the science detector, for example, causing the PSF of the star to elongate in a particular direction or show variable elongations in various directions\cite{2016SPIEGemini}. The effect of vibrations can impact the quality of the calibrations, and consequently, the performance of the system itself. 

It is, therefore, essential to know the vibration environment of the telescope and to predict, dampen, and mitigate the vibrations as much as possible. In almost all modern telescopes, there are vibration measurement and compensating systems. There is a vibration measurement system in the LBT called the Optical Vibration Monitoring System (OVMS)\cite{2010SPIEOVMSMK}. It uses a network of accelerometers to monitor vibrations on the telescope. Measurements from the updated version of OVMS, called OVMS-plus (or OVMS+)\cite{2016SPIEOVMSplusMB}, could track down the causes of vibrations on the telescope in various observing conditions\cite{2016SPIEHill}. The recent updates about the vibration environment at LBT can be seen in the Esc\'arate et al. paper\cite{2018SPIEPedro}. In the past few months, the LBT is implementing the OVMS+ tip and tilt signals in the AO control loop in a disturbance feed-forward manner to mitigate the vibrations\cite{Bohm:17}.

\begin{figure}[htbp]
\centering
\fbox{\includegraphics[width=0.9\linewidth]{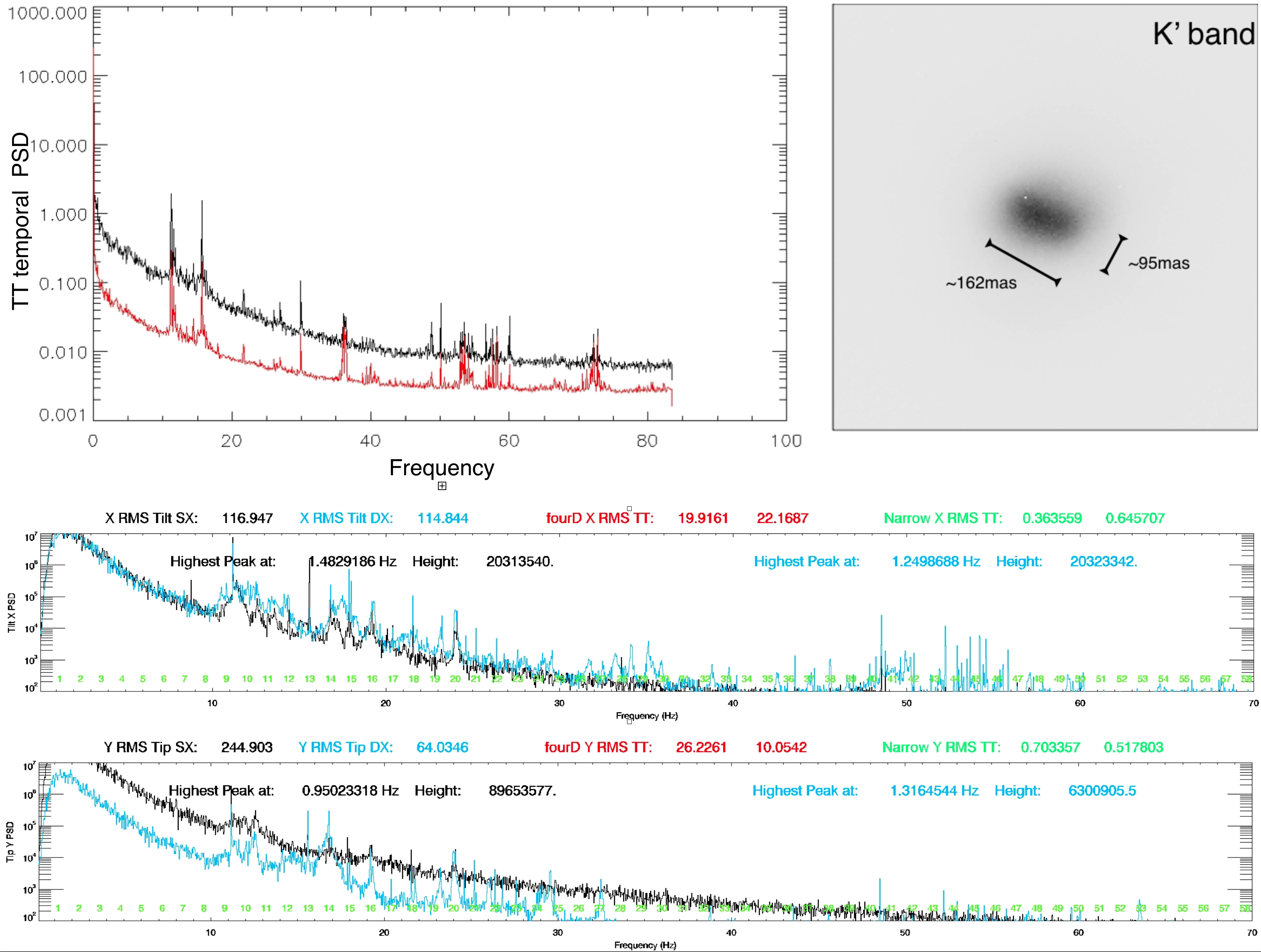}}
\caption{An example of vibrations seen by LN. \textit{Top-left} - The power spectral density of the vibrations from the GWS WFS data normalised to 1. \textit{Top-right} - Cigar shaped PSF seen by the LN science camera while the ground layer loop was closed. \textit{Bottom} panels show the vibration measured by the OVMS+ during the same night.}
\label{fig1}
\end{figure}

We have observed relatively strong vibrations at 9\hspace{1pt}Hz and 16\hspace{1pt}Hz on a few commissioning nights in December 2018. The image on the top-right of Figure~\ref{fig1} displays the PSF of a star in the K' band when these vibrations were severe. While this image was captured by the LN scientific camera, the ground-layer turbulence was removed by the GWS (correcting up to 30 KL modes). The shape of the PSF appears like the shape of a cigar, with the FWHM of the longer side almost double that of the shorter side. Although diminished by the AO system, the residual vibrations are limiting the performance of the system and therefore the PSF at the science detector. Note that visually (and later using the saved images) we noticed that the peak of the PSF improved by a factor of $\sim$2 at the science detector with the GWS loop closed, implying a very good correction by the GWS. OVMS+ also detected the peaks at the same frequencies, displayed in the bottom panels of Figure~\ref{fig1}.

The exact reasons causing these vibrations are still unknown. Switching on and off different possible candidates within LN and the LBT structure have so far not provided any conclusive evidence of the source of these vibrations. In addition, these vibrations are not seen on an every night basis. However, measures are ongoing to find the source, and minimize or remove it. It is worth pointing out that other AO-fed instruments at the LBT also see similar vibrations occasionally. 

\section{AO Calibration}
The quality of the Interaction Matrix (IM) of an AO system is one of the determining factors of its performance. Therefore, AO calibration plays a vital role in the working of any AO-fed instrument. LN has an optical/mechanical derotator in between the DM and the wavefront sensor and thus a continually changing actuator to wavefront sensor sub-aperture relationship as the sky rotates during observation. The calibration procedure\cite{2018SPIECarmelo,2010SPIECarmelo} can take between 3-6 days (8 hours a day) to measure and create good quality (low condition number) IMs at different derotation angles. 

The interaction matrix calibrations are performed with a closed dome during day time with extremely quiet conditions. However, there are always some vibrations. In order to avoid the effects of the existing vibrations spread across the entire interaction matrix measurements, we follow this strategy. First, only the tip and tilt modes are applied on the deformable mirror (with rather high amplitudes but while not saturating the WFS and in its linear range) and measure the slopes at the WFS and create the interaction matrix with only for these two modes. Later, the measurements are made for higher number of modes. Finally, the tip and tilt are measured with an amplitude consistent to the one used for the other modes, taking advantage of the high order mirror flat produced by closing high-order loop. From the high-order measurements, the tip-tilt component is filtered out and adjoining the interaction matrix for only the tip and tilt modes. We have tested our calibration procedure and verified on-sky.

Recently, the team created and tested pseudo-synthetic IMs. Of course, these IMs also need to be tuned to the various parameters of the system. This means that one set of good calibrations for a known derotation angle is essential. We found that the condition numbers of the pseudo-synthetic IMs were lower than the measured ones and proved to be significantly more efficient in closing the loop. Since computing power is not an issue these days, pseudo-synthetic IMs shows promise for upcoming MCAO systems.

\section{Operational Related}
\subsection{Temperature Variations at LBT}
The average night temperature at the LBT generally ranges between -20\hspace{1pt}\textdegree C and +20\hspace{1pt}\textdegree C \cite{2016SPIEHill} with a maximum observed daytime temperature of 28\hspace{1pt}\textdegree C. The broad range of temperatures has implications on the general operations of the instrument, calibrations, collimation, etc. The cooling within the electronic cabinets and to the detectors need to be controlled accordingly. For example, during one of the runs in the winter time, the fans had to be switched off so that the CCD controllers were sufficiently warmed up. In contrast, during the summer of 2017 in the daytime, the cabinets were automatically turned off after half an hour of switching on the electronics, even with the maximum cooling flow rate. Perhaps, it may be a good idea to have thermal enclosures for at least part of the instruments where the optics has high impact on the high range of temperature fluctuations. For example, the response of the actuators (or the influence function) of the deformable mirrors are usually valid within a specific temperature range.

\subsection{Acquisition}
As a multi-pyramid system, it is essential to have the acquisition of both the ground-layer and the high-layer natural guide stars as quickly as possible to minimise the initial overhead. The two main factors are (1)~the accuracy of star charts and instrument characterization and (2)~the search and centering software. The better the accuracy of the star positions (both from catalogues and at the acquisition software), the less time spent searching for photons. Note that, due to the design of the optics, the FoV at each of the pyramid is 1.1\hspace{1pt}" in diameter. Once there are photons in the pupil(s), the software has to be efficient in centering the probe before moving on to the next star. Note that the search and centering algorithm must be robust to changing conditions (different magnitude stars, varying seeing conditions, etc.). The light detection is a balancing act between robustness and accuracy (avoid false detection but also detect stars barely above noise threshold). For the first point, Gaia catalogues\cite{GaiaDR1,GaiaDR2} are of immense help. Our acquisition software has substantially improved over the last two years. At this point, we are able to acquire 4 ground-layer stars (of magnitude close to 11 in R-band) and 3 high-layer stars (of magnitude close to 11 in R-band) in 3 minutes. We were also able to acquire and center stars in the range 13-15\hspace{1pt}mag (R-band) with slower frame-rate and higher binning.

\subsection{Flexure Tracking}
LBT is an alt-azimuth telescope. Therefore, the instruments located on the LBT platform are not in a gravity invariant position. Flexure tracking is crucial, especially for consistent and reliable AO performance. For example, the LN WFS CCDs drift according to the elevation of the telescope. In order to compensate for this drift, the team has implemented a CCD tracking software using the optical input from the WFS CCDs. Although this may appear easy, for fainter guide stars, maintaining the misregistration between the actuator map and the WFS map less than one-tenth of a sub-aperture is not trivial. Note that the WFS CCDs have to maintain their position not only in lateral XY axes but also in the focus Z axis through out the observation in real-time.

\subsection{Offsetting the Telescope}
For infra-red observations, it is common to offset the telescope by a few arcseconds to arcminutes to measure the sky background. In some cases, we need to do this with the AO loop closed. A typical example would be if the observer requires mosaic frames. In the case of LN, we have a certain range of motion for the probes acquiring the natural guide stars. This is possible as long as the movement is within their ranges and in favourable directions. We have tested this, with a closed ground loop, by moving the acquired probes uniformly in the same direction. However, the maximum step size is $\sim$0.8\hspace{1pt}" in order to guarantee that the ground loop will remain closed. Therefore, this procedure may be ideal for small FoV adjustments or dithering.

\subsection{Maintenance and Working on the LN Bench}
Even after the integration of the instrument at the telescope, there are times when we need to go to the optical bench during the commissioning phase of an instrument. This may be necessary to investigate if all the optical components and their mechanical parts are behaving well, or if there is something vignetting the FoV, or to make small changes or tweaks, or in an unfortunate case of misalignment of the optics to make the proper corrections, etc. We have gone through similar situations during the LN commissioning. Working on the LN bench at the telescope is different from working at the lab, especially during winter. In addition, you do not always have the same degrees of freedom to reach or move the optics. The take away message from here is that while designing the optics, the size of the bench, etc., it is worth considering the people who will actually be aligning them and performing the maintenance. 

\subsection{CCD39 Aurora Effect}
\begin{figure}[htbp]
\centering
\fbox{\includegraphics[width=0.95\linewidth]{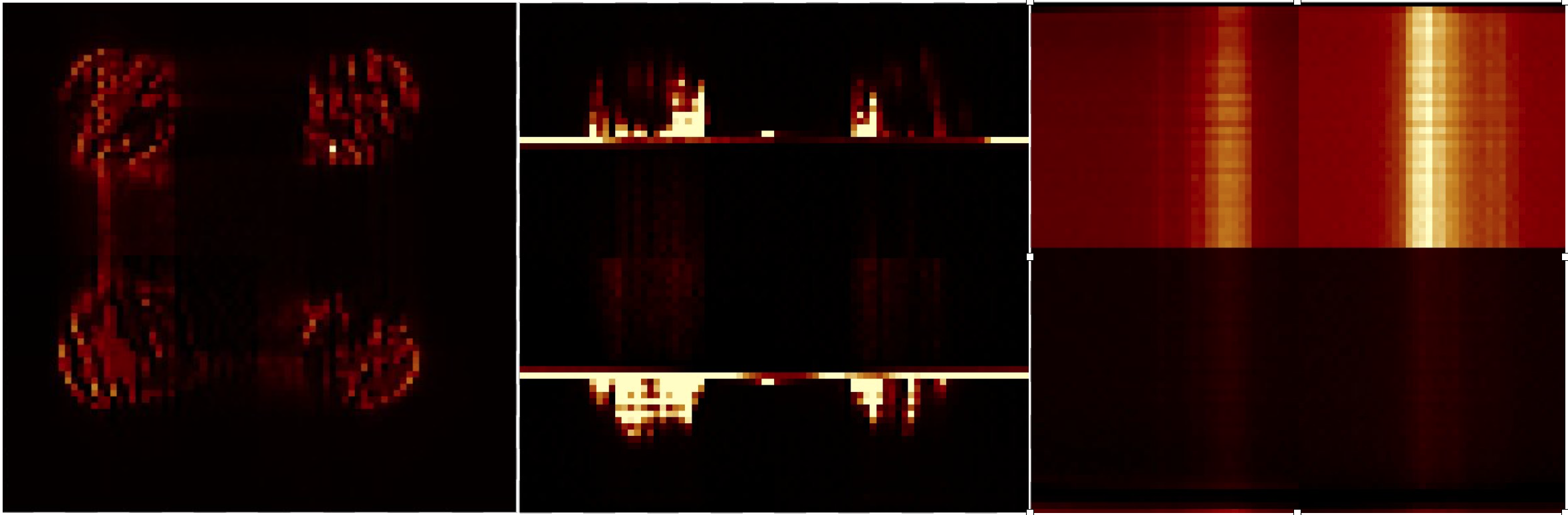}}
\caption{CCD39 Aurora Effect. The images from left to right show the various stages, i.e. the good signal, the intermediate unstable signal, and the complete aurora effect signal or ``the good, the bad, and the ugly".}
\label{fig2}
\end{figure}

During one of the pre-commissioning runs in November 2016, we encountered a strange issue with the high-layer WFS CCD39. Positioning a probe at the on-axis star (a light source), we could not see an image of the 4 pupils on the detector, but only a strange light pattern (showing that light is actually reaching the 4 quadrants of the CCD, but CCD is not showing pupils). We tried with a different probe and we had the same problem. This behaviour was shown by both the CCD39 detectors on either sides. Notably, it was working fine the day before. We call this the ``aurora effect," since the pattern resembles Aurora Borealis (see Figure \ref{fig2}).

We tested various possibilities one by one to rule them out as the cause. Finally, we found the issue. It was a temperature problem. The CCD39 electronics are installed inside the cabinets. The cabinets are continuously temperature controlled. Switching off the fan and the heat exchange units, the signals became better. Switching off the cabinet cooling made the signal as it should be. This was reproducible as well, meaning switching on the cooling of the cabinets, the fan and the heat exchange units the aurora effect returned. With the help of cold spray the problematic ROE module was identified to be the support module. The support module is in the last slot of the back-plane and, in this spot, exposed to efficient convection. The idea was to shield the module from the air flow of the fans that are located right below the rack with the CCD39 electronics. A piece of cardboard was installed and the area below the support module was blocked with aluminium tape. This worked and the aurora effect was solved.

\subsection{Splitting the Commissioning Nights}
At the time of the writing of this paper, the LN team has made 9 commissioning runs over two and a half years. The first 6 runs consisted of primarily half-nights for nighttime commissioning tasks. This was preferred by the team to also accomplish daytime tasks. Also, in the initial runs, the learning curves in several departments were steep. We had to fix the errors and bugs we found during the nighttime. This approach not only helped the commissioning team but also saved precious night telescope time. There were situations when we were stuck during the end of the first-half of the night and the second-half observing team were happy to take over before their actual time.
\begin{figure*}[htbp]
\centering
\fbox{\includegraphics[width=0.95\linewidth]{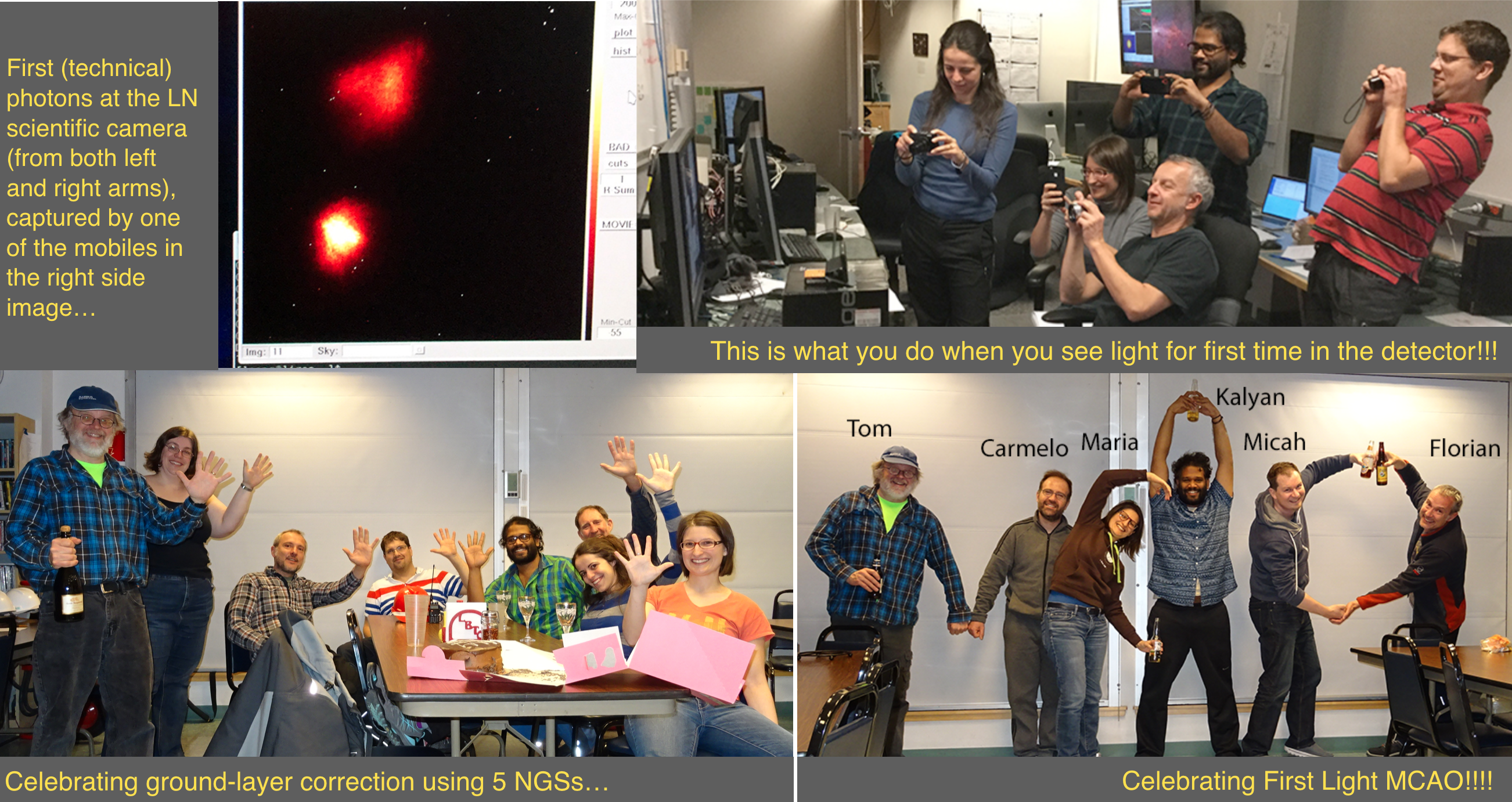}}
\caption{\textit{Top-left} - First technical photons reaching the LN science detector. \textit{Top-right} - LN team capturing the first technical photon moment in their mobiles. \textit{Bottom-left} - Team celebrating the successful ground-layer correction using 5 NGSs. \textit{Bottom-left} - Team celebrating the LN First Light MCAO.}
\label{fig3}
\end{figure*}

\acknowledgments
The authors express their sincere gratitude to the LBT mountain crew for their continuing support and dedication, which made our remote and on-site activities very smooth and effective.

The LBT is an international collaboration among institutions in the United States, Italy, and Germany. LBT Corporation partners are: The University of Arizona on behalf of the Arizona university system; Istituto Nazionale di Astrofisica, Italy; LBT Beteiligungsgesellschaft, Germany, representing the Max-Planck Society, the Astrophysical Institute Potsdam, and Heidelberg University; The Ohio State University, and The Research Corporation, on behalf of The University of Notre Dame, University of Minnesota, and University of Virginia. 

The authors also acknowledge and thank the continuing support provided both on-site and remote by all the engineers and support staff within the LN consortium.

\bibliography{report} 
\bibliographystyle{spiejour} 

\end{document}